\documentclass[referee]{raa}
\usepackage{graphicx,times}             
\usepackage{natbib}
\usepackage{amssymb,amsmath}
\usepackage{multirow}
\usepackage{booktabs}
\usepackage{color}
\usepackage{ulem}
\bibpunct{(}{)}{;}{a}{}{,}
\usepackage[a4paper=true,dvipdfm=true,pagebackref=true]{hyperref}
\hypersetup{colorlinks = true, linkcolor = green, anchorcolor = red, citecolor = blue, filecolor = red, pagecolor = red, urlcolor = red}

\begin{document}
 \title{Site-testing at Muztagh-ata site I: Ground Meteorology and Sky Brightness
}

   \volnopage{Vol.0 (20xx) No.0, 000--000}      
   \setcounter{page}{1}          

   \author{Jing Xu
      \inst{1,2}
   \and Ali Esamdin
      \inst{1,2}
   \and Jin-xin Hao
      \inst{3}
   \and Jin-min Bai
      \inst{4}
   \and Ji Yang
      \inst{5}
   \and Xu Zhou
      \inst{3}
   \and Yong-qiang Yao
      \inst{3}
   \and Jin-liang Hou
      \inst{6}
   \and Guang-xin Pu
      \inst{1}
   \and Guo-jie Feng
      \inst{1,2}
   \and Chun-hai Bai
      \inst{1}
   \and Peng Wei
      \inst{1}
   \and Shu-guo Ma
      \inst{1}
   \and Abudusaimaitijiang Yisikandee
      \inst{1}
   \and Le-tian Wang
      \inst{1}
   \and Xuan Zhang
      \inst{1}
   \and Liang Ming
      \inst{1}
   \and Lu Ma
      \inst{1}
   \and Jin-zhong Liu
      \inst{1}
   \and Zi-huang Cao
      \inst{2,3}
   \and Yong-heng Zhao
      \inst{3}
   \and Lu Feng
      \inst{3}
   \and Jian-rong Shi
      \inst{3}
   \and Hua-lin Chen
      \inst{7}
   \and Chong Pei
      \inst{7}
   \and Xiao-jun Jiang
      \inst{3}
   \and Jian-feng Wang
      \inst{3}
   \and Jian-feng Tian
      \inst{3}
   \and Yan-jie Xue
      \inst{3}
   \and Jing-yao Hu
      \inst{3}
   \and Yun-ying Jiang
      \inst{3}
}

   \institute{Xinjinag Astronomical Observatory, Chinese Acsdemy of Sciences, Urumqi, 830011, People's Republic of China; {\it xujing@xao.ac.cn,aliyi@xao.ac.cn}\\
        \and  University of Chinese Academy of Sciences,
             Beijing 100049, People's Republic of China\\
        \and
             National Astronomical Observatories, Chinese Academy of Sciences,
             Beijing 100012, People's Republic of China\\
        \and
             Yunnan Observatories, Chinese Academy of Sciences,
             Kunming 650000, People's Republic of China\\
        \and
            Purple Mountain Observatories,Chinese Academy of Sciences,
            Nanjing 210008, People's Republic of China\\
        \and
            Shanghai Astronomical Observatories, Chinese Academy of Sciences,
            Shanghai 200030, People's Republic of China\\
        \and
            National Astronomical Observatories Nanjing Institute of Astronomical Optics \& Technology,
            Chinese Academy of Sciences,
            Nanjing 210008, People's Republic of China\\
\vs\no
   {\small Received~~20xx month day; accepted~~20xx~~month day}}

\abstract{ Site-testing is crucial for achieving the goal of scientific research and analysis of meteorological and optical observing conditions is one of the basic tasks of it. As one of three potential sites to host 12-meter Large Optical/infrared Telescope (LOT), Muztagh-ata site which is located on the Pamirs Plateau in west China's Xinjiang began its site-testing task in the spring of 2017. In this paper, we firstly start with an introduction to the site and then present a statistical analysis of the ground-level meteorological properties such as air temperature, barometric pressure, relative humidity, wind speed and direction, recorded by automatic weather station with standard meteorological sensors for two-year long. We also show the monitoring results of sky brightness during this period.
}
   \authorrunning{J. Xu et al}            
   \titlerunning{Site-testing at Muztagh-ata site I: Ground Meteorology and Sky Brightness }  

   \maketitle

%
\section{Introduction}           
\label{sect:intro}
The international astronomical community has sustained great effort to find the best sites for the operation of large optical telescopes over the last few decades. An excellent site is the crucial condition for constructing the observational facilities of which the performance was determined by it. Recent year, the projects of new-generation astronomical facilities such as the Thirty Meter Telescope (TMT)\citep{2009PASP..121..384S}and the European Extremely Large Telescope (E-ELT)\citep{2012MNRAS.422.2262R} have been committed to a comprehensive program of site selection and promoted the theory in evaluation.

With the rapidly growing of astronomy in China recent years, an ambitious project, the Large Optical/infrared Telescope (LOT), with a goal to construct a 12-meter telescope was elected to be built in the next decade. In order to maximize the performance of the telescope, A site assessment campaign has been initiated for more than two years to identify the most suitable location to host it \citep{Feng2019}.

In the first decade of this century, \cite{2005JKAS...38..113Y} surveyed the astronomical resource of west China and suggested that the best candidate sites may be located in the eastern Pamirs. Through the satellite photographs and field survey on the Pamirs plateau, the Muztagh-ata site was selected as one of the three alternatives, the other two are Shiquanhe site in Tibet\citep{Liu2019} and Daocheng site in Sichuan province\citep{Song2019}.
The gathering of meteorological data is essential prior to any investment in equipment or telescopes. The properties of meteorological should be considered in site selection include air temperature, relative humidity, wind speed and wind gust, etc\citep{2002ASPC..266..498M, 2000A&AS..147..271J, 2016PASP..128j5003T}.
The best observing sites should have small daytime and nighttime air temperature gradients,high stability of the air temperature during the night, small relative humidity, and low wind speed and wind gusts\citep{2014PASP..126..412V, 1998NewAR..42..417M, 1985VA.....28..449M}.
Although the monitor data available only for two year cannot provide the conclusion of climate in this area, statistical analysis of meteorological properties is available for the LOT campaign. Long-term analysis of cloudy amounts through satellite data, data processing method and results of all-sky images for this area were concluded in \citep{Cao2019a,Cao2019b}, during the period from 10th March 2017 to 10th March 2019, ``clear" (No cloud) and ``outer" (No cloud within the inner circle while some cloud within the outer circle of all-sky images) account for 61.6$\%$ of nighttime and 63.1$\%$ of nights ``clear" + ``outer" more than three hours.

The layout of this paper is as follows: In Section 2, we describe the Muztagh-ata site from the aspect of topography, traffic, surrounding social situation, etc. In Section 3, we detail the monitoring instruments. In Section 4, we show the statistical results of site monitoring during the two years, which include air temperature, relative humidity, barometric pressure, wind speed and direction and sky brightness respectively. In Section 5, we summarize the final conclusions.
\section{Site Description }
\begin{figure}
   \centering
  \includegraphics[width=8cm, angle=0]{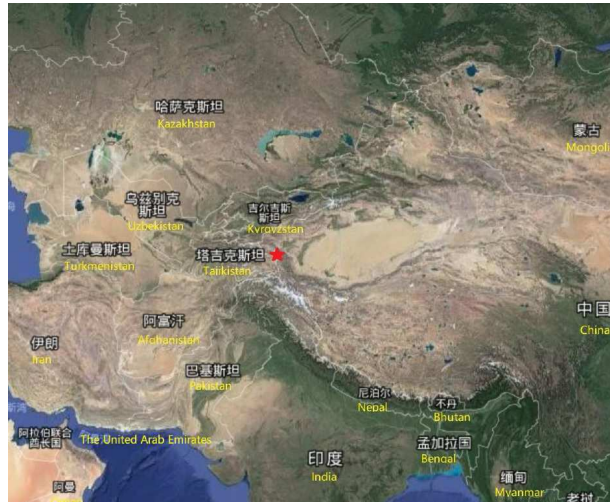}
   \caption{The red pentagram in this map represents the location of Muztagh-ata site, geographical location is 38$^{\circ}$19'47''N, 74$^{\circ}$53'48''E, in southwest of Xinjiang Uygur Autonomous Region of China.}
   \label{fig:xj_1}
\end{figure}

The Muztagh-ata site is located in the eastern part of the Pamirs Plateau, in southwest of Xinjiang Uygur Autonomous Region of China. The geographical coordinates are 38$^{\circ}$19'47''N, 74$^{\circ}$53'48''E, with an altitude 4526 meters above sea level. The site locations are given in the map\footnote{https://map.baidu.com/} in Figure~\ref{fig:xj_1}. The Muztagh-ata, 7549-meter-high, lies to the east of it, so we refer to our site as the Muztagh-ata site.

Our site lies to the southwest of the city of Kashgar, which is the third largest city of Xinjiang and has multiple good transport links. The site can be accessed by 200 $km$ national road from Kashgar.
There are several high mountains taller than 5000 $meters$ surrounding the site in a range of 100 $km$, blocking the Indian Ocean wet airflow and the dust of the Tarim Basin from going to the site, which provide a relative stable weather conditions for this area. The local climate shows a typical plateau continental characteristic with an oxygen deficit, cold, dry, and receives little precipitation. According to the statistics provided by the local meteorological department the local meteorological conditions were described as follows. ¡°The oxygen deficit is 30$\%$ $\sim$ 40$\%$. The annual average temperature is -6 $^{\circ}$$C$. The extreme low temperature recorded is -36 $^{\circ}$$C$. The average annual precipitation is 271.1 $mm$ and solid precipitation predominates, most in the late spring and summer. Southwest wind is prevailing wind direction at the surface level.¡± These data indicates the existence of potentially good sites in this area for optical observations.

The light pollution situation is very good: Our site is in a sparsely populated region where the herdsmen live for a very short time in summer. The national road passes by the foot of the mountain but the relative altitude between it and the site is $\sim $900 $m$.

\begin{figure}
   \centering
  \includegraphics[width=8cm, angle=0]{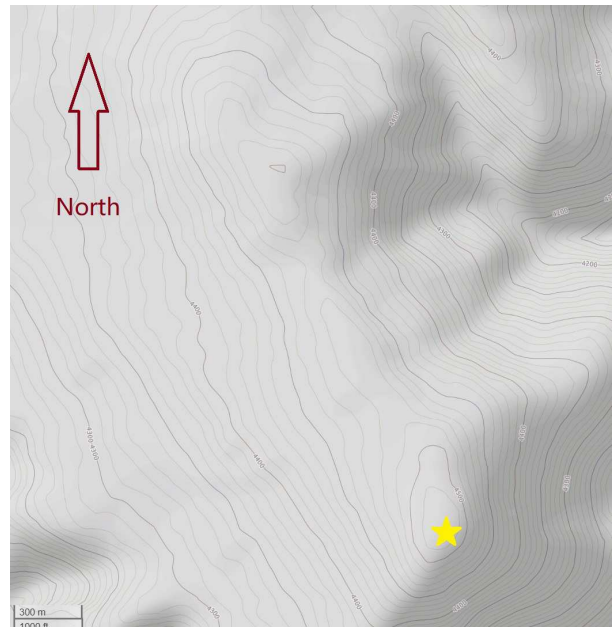}
   \caption{Contour map of the region surrounding our site. The yellow pentagram represents the location of our monitoring equipments at present.}
   \label{fig:xj_2}
\end{figure}

\begin{figure}
   \centering
  \includegraphics[width=8cm, angle=0]{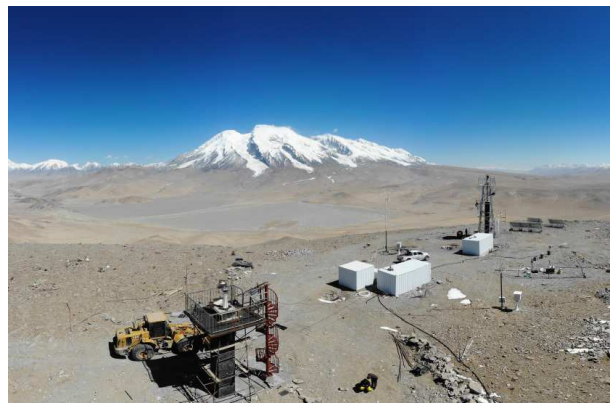}
   \caption{General view of Muztagh-ata site.The snowmountain lies in the background is Muztagh-ata}
   \label{fig:xj_3}
\end{figure}

In Figure~\ref{fig:xj_2} a contour map\footnote{https://www.openstreetmap.org/\#=13/38.3473/74.9173\&layers=C} shows the topographic conditions surround Muztagh-ata site, there is a piece of $\sim $1 $km$$^2$ flat region on the top of the mountain can be available for accommodating the infrastructure for a large telescope. The yellow pentagram in Figure~\ref{fig:xj_2} represents the location of our monitoring point at present, so far we have built two towers for the Differential image motion monitor (DIMM) and some concrete foundations for various monitoring facilities there. Figure~\ref{fig:xj_3} is the general view of Muztagh-ata site.

\section{Monitoring Instruments- Auto Weather Station}
Climatic parameters of our site are derived from the data of the auto weather station, which is a five-element commercial high-precision meteorological station. It measures and records air temperature, barometric pressure, relative humidity, wind speed and wind direction values. The anemometer and wind vane were supported by a 10-meter tower. The air-temperature sensor, the hygrometer and the barometer were set in a thermometer screen 1.5 meters from ground level. All of those sensors are calibrated to the specified absolute measurement accuracies in Table~\ref{Tab:xj_1}. The equipment is powered by solar panels and transmitting data through GSM signal to the server hosted at National Astronomical Observatory the Chinese Academy of Sciences (NAOC) every 10 minutes and the sampling interval is one minute.

The auto weather station was installed on March 2$^{nd}$, 2017. There were several days data records absence in the two-year site-testing because of the disappearance of GSM signal caused by the power failure of the communication base stations. In addition, due to sensors failure, the recorded periods of each parameter vary, showed in Table~\ref{Tab:xj_1} last column.

\begin{table*}
\centering
\caption{Ranges Accuracies and Recorded Periods of Meteorological Sensors.}
\label{Tab:xj_1}
\begin{tabular}{llll} 
\hline
Element & Range & Accuracy&Record period \\
\hline
Air temperature/$^{\circ}$$C$&-50$\sim$50&$\pm$0.1&2$^{nd}$ Mar. 2017$\sim$28$^{th}$ Feb. 2019\\
Relative humidity/RH$\%$&0$\sim$100&$\pm$3&1$^{th}$ Aug. 2017$\sim$ 31$^{st}$ Jul. 2019\\
Wind speed/$ms$$^{-1}$&0.5$\sim$50&$\pm$0.5&2$^{nd}$ Mar. 2017$\sim$ 28$^{th}$ Feb. 2019\\
Wind direction/$^{\circ}$&0$\sim$360&$\pm$6&12$^{th}$ Mar. 2017$\sim$ 28$^{th}$ Feb. 2019\\
Barometric pressure/$hPa$&300$\sim$1100&$\pm$1&2$^{nd}$ Mar. 2017$\sim$ 28$^{th}$ Feb. 2019\\
\hline
\end{tabular}
\end{table*}

\section{Site Monitoring Results}
\subsection{Air Temperature}
\begin{figure}
   \centering
  \includegraphics[width=12.5cm, angle=0]{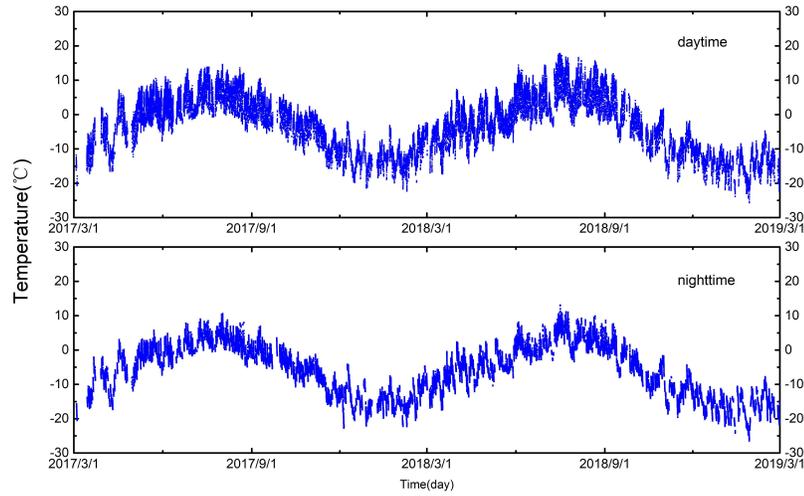}
   \caption{Air temperature values with time over daytime (above) and nighttime (below) periods.}
   \label{fig:xj_4}
\end{figure}

\begin{figure}
   \centering
  \includegraphics[width=12.5cm, angle=0]{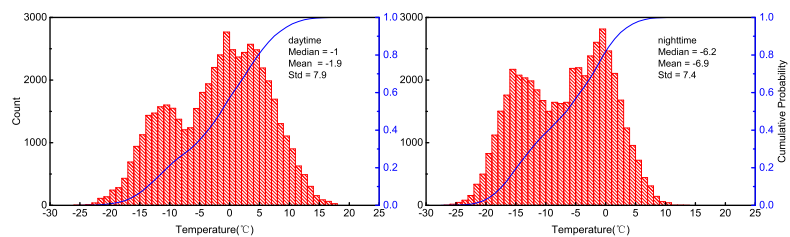}
   \caption{Histograms and cumulative distributions of air temperature over daytime (left) and nighttime (right) periods.}
   \label{fig:xj_5}
\end{figure}

We analyze the differences of all meteorological parameters between the day and night, for this purpose we define the time from the beginning of astronomical morning twilight to the end of astronomical evening twilight as daytime, and the rest of this day as nighttime.
Figure~\ref{fig:xj_4} shows the daytime and nighttime air temperature with time at Muztagh-ata for the whole dataset during the measurement period indicated in Table~\ref{Tab:xj_1}. A strong seasonal dependency can be seen from this figure. The distributions and cumulative statistics for daytime and nighttime are shown in Figure~\ref{fig:xj_5}, the mean value of the daytime data is -1.9$^{\circ}$$C$, the median value is -1.0$^{\circ}$$C$. The mean value of the nighttime data is -6.9$^{\circ}$$C$, the median value is -6.2$^{\circ}$$C$. There are two peaks in each histogram in Figure~\ref{fig:xj_5}, represent the median temperature values in summer and winter respectively.

\begin{figure}
   \centering
  \includegraphics[width=13.5cm, angle=0]{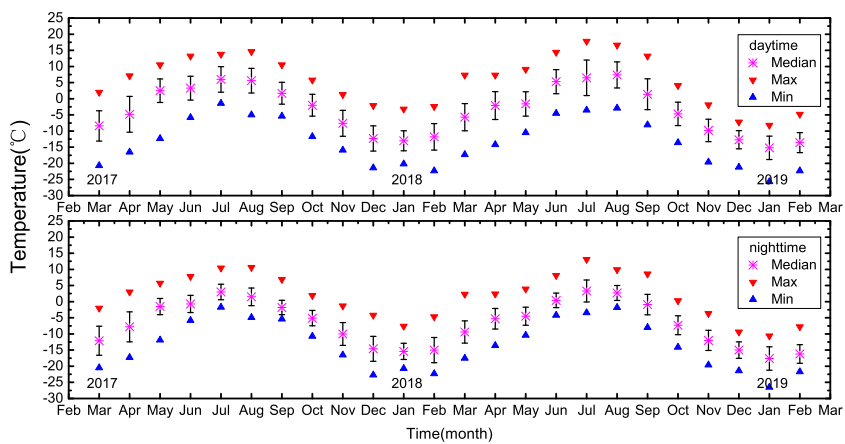}
   \caption{Monthly statistic of the air temperature over daytime (above) and nighttime (below) periods from March 2017 to February 2019.}
   \label{fig:xj_6}
\end{figure}

We calculate the median, maximum and minimum values of the temperature per month from March 2017 to February 2019 over daytime and nighttime periods separately represented in Figure~\ref{fig:xj_6}. Error bars correspond to the standard deviation of the mean values. Detail monthly statistics results were shown in Table~\ref{Tab:xj_2}, January is the coldest month in this area with daytime mean value of -14.1$^{\circ}$$C$ and nighttime mean value of -16.4$^{\circ}$$C$ while the June is the warmest month with daytime mean value of 6.3$^{\circ}$$C$ and nighttime mean value of 3.2$^{\circ}$$C$. The maximum value (17.9$^{\circ}$$C$) occurred on 16$^{th}$ July 2018 and the minimum value (-26.5$^{\circ}$$C$) occurred on 28$^{th}$ January 2019.
\begin{figure}
   \centering
  \includegraphics[width=12.5cm, angle=0]{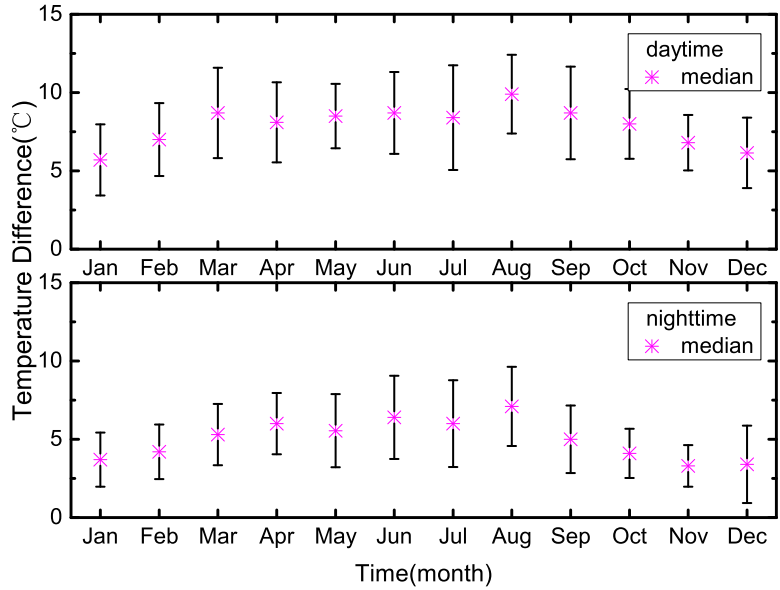}
   \caption{Monthly median and standard deviation values over daytime (above) and nighttime (below) periods of temperature difference.}
   \label{fig:xj_7}
\end{figure}

In order to explore the dependency of temperature differences during daytime and nighttime on season we calculate the daily median and standard deviation values of the temperature differences during daytime and nighttime. Then we show the monthly statistics results in Figure~\ref{fig:xj_7}. The median temperature difference values are higher in warm seasons than in cool seasons. The daily daytime temperature difference median value of August is 9.9$^{\circ}$$C$ and nighttime is 7.1$^{\circ}$$C$.

\begin{table*}
\centering
\caption{Monthly statistics on Daytime and Nighttime Data of air temperature.}
\label{Tab:xj_2}
\begin{tabular}{lllllllllll}
\hline
\multirow{2}{*}{Month} & \multicolumn{5}{c}{Nighttime Temperature} &\multicolumn{5}{c}{Daytime Temperature}\\ \cmidrule(lr){2-6} \cmidrule(lr){7-11}
   &Max &Min  &Median  &Mean  &Std &Max &Min  &Median  &Mean  &Std \\
\hline
Jan. &-7.6& -26.5& -16.0& -16.4& 3.4& -3.2& -25.7& -13.7& -14.1& 3.7\\
Feb. &-4.7& -22.3& -15.6& -15.0& 3.6& -2.4& -22.3& -12.7& -12.3& 3.8\\
Mar. &2.3& -20.5& -10.3& -10.4& 3.9& 7.3& -20.7& -5.9& -7.0& 4.5\\
Apr. &3.0& -17.3& -6.4& -6.7& 4.1& 7.3& -16.5& -3.1& -3.5& 5.1\\
May. &5.7& -11.9& -1.7& -1.5& 2.5& 10.5& -12.3& 2.0& 2.5& 3.6\\
Jun. &8.1& -5.8& -0.2& 0.0& 2.6& 14.4& -5.8& 4.0& 3.9& 3.8\\
Jul. &13.1& -3.4& 3.1& 3.2& 3.0& 17.8& -3.5& 6.2& 6.3& 4.8\\
Aug. &10.6& -4.9& 2.1& 2.2& 2.6& 16.6& -5.0& 6.4& 6.2& 4.0\\
Sep. &8.6& -8.0& -1.3& -1.3& 2.8& 13.2& -8.1& 1.6& 1.7& 4.2\\
Oct. &1.9& -14.1& -5.9& -6.1& 2.9& 5.8& -13.6& -3.1& -3.3& 3.7\\
Nov. &-1.3& -19.6& -11.2& -11.0& 3.5& 1.3& -19.6& -8.8& -8.6& 4.0\\
Dec. &-4.2& -22.7& -14.9& -14.8& 3.3& -2.1& -21.4& -12.5& -12.5& 3.4\\
Total &13.1& -26.5& -6.2& -6.9& 7.4& 17.8& -25.7& -1.0& -1.9& 7.9\\
\hline
\end{tabular}
\end{table*}

\subsection{Relative Humidity}
\begin{figure}
   \centering
  \includegraphics[width=12.5cm, angle=0]{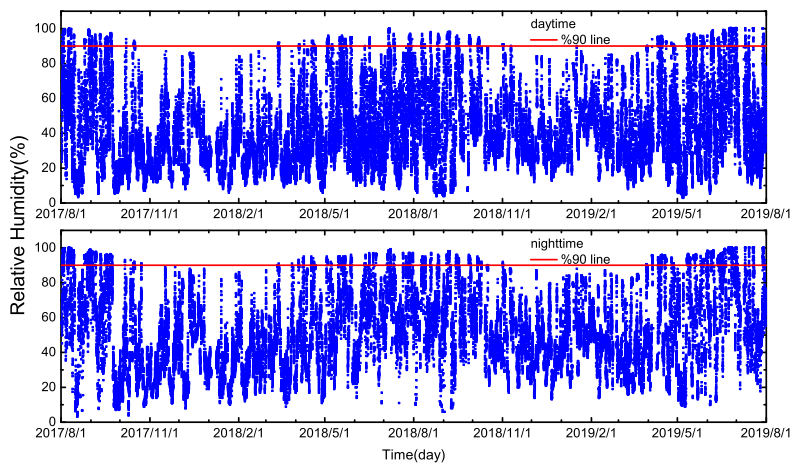}
   \caption{Relative humidity with time over daytime (above) and nighttime (below) periods from August 2017 to July 2019 and 90$\%$ lines.}
   \label{fig:xj_8}
\end{figure}

\begin{figure}
   \centering
  \includegraphics[width=12.5cm, angle=0]{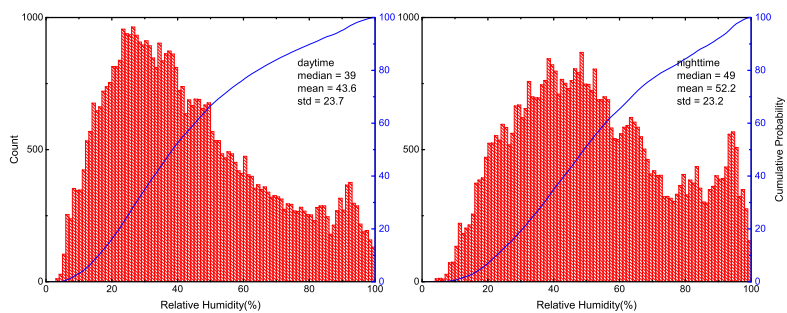}
   \caption{Relative humidity histograms and cumulative distributions over daytime (left) and nighttime (right) periods.}
   \label{fig:xj_9}
\end{figure}

\begin{figure}
   \centering
  \includegraphics[width=12.5cm, angle=0]{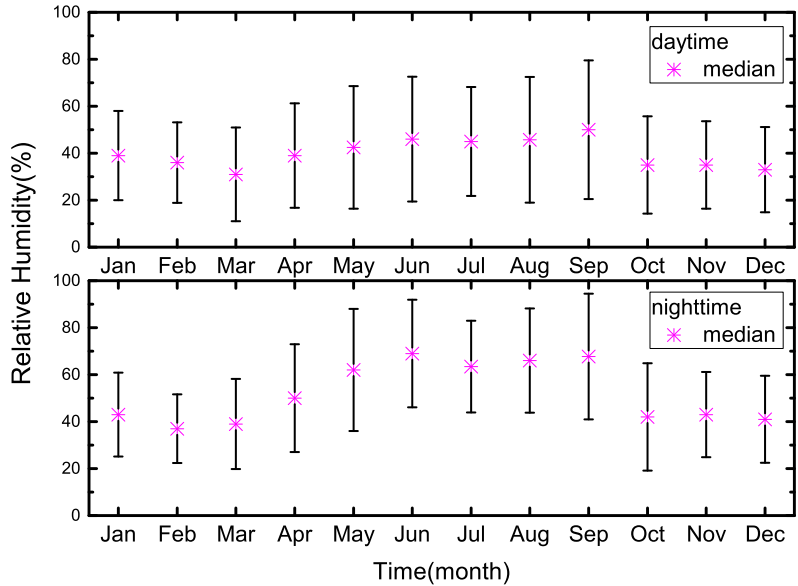}
   \caption{Monthly median and standard deviation values over daytime (above) and nighttime (below) periods of relative humidity.}
   \label{fig:xj_10}
\end{figure}

Because of sensors failure, relative humidity data is available only from August 2017, to July 2019 twenty four months in total. Figure~\ref{fig:xj_8} shows the daytime and nighttime relative humidity with time during this period.
The relative humidity and the dew point are two important parameters for the astronomical instrumentation, because they set the occurrence of moist and water condensation on the coldest parts of astronomical instrumentation, which determines whether the observation can be made\citep{2009MNRAS.399..783L}.
If the humidity is high enough, or when the dew point temperature is very close to the air temperature, condensation will easily occur\citep{1985VA.....28..449M}.
Red lines in Figure~\ref{fig:xj_8} are 90$\%$ lines as the threshold, beyond which the observations should be stopped, there are 6876 of the 95817 data points of which the values are higher than 90$\%$, account for 7.2$\%$, the rate is 8.8$\%$ of nighttime data.
The distributions and cumulative statistics for daytime and nighttime of relative humidity over daytime and nighttime were represented in Figure~\ref{fig:xj_9}, the median value of daytime is 39$\%$ and 49$\%$ for nighttime respectively. Monthly statistics results over daytime and nighttime are shown in Figure~\ref{fig:xj_10}, some detailed information of monthly statistics is shown in Table 3. The highest monthly average of nighttime data is 68.7$\%$ , in June. The relative humidity is higher in late summer and autumn of each year.

\begin{table*}
\centering
\caption{Monthly statistics on Daytime and Nighttime Data of relative humidity.}
\label{Tab:xj_3}
\begin{tabular}{lllllllllll}
\hline
\multirow{2}{*}{Month}
        & \multicolumn{5}{c}{Nighttime Relative Humidity} &\multicolumn{5}{c}{Daytime Relative Humidity}\\
         \cmidrule(l){2-6} \cmidrule(l){7-11}
    &Max &Min  &Median  &Mean  &Std &Max &Min  &Median  &Mean  &Std \\
\hline
Jan. &88	&10&43&44.7&17.9&90&6&39&39.9&19.0\\
Feb. &87&12&43&43.2&15.7&87&10&36&37.9&17.1\\
Mar. &91&9&39&43.3 &19.2 &92&8&31&36.6&20.0 \\
Apr. &96&11&50&55.0 &23.0 &96&6&39&43.7&22.3 \\
May. &97&9&62&58.6 &26.0 &97&3&43&47.3 &26.1 \\
Jun. &100&10&69&68.7 &22.9 &100&5&46&50.4 &26.6 \\
Jul. &100&11&63&66.2 &19.5 &100&8&45&47.5&23.2 \\
Aug. &100&3&66&64.7 &22.2 &100&3&46&46.3 &26.8 \\
Sep. &98&6&68&63.7 &26.8 &98&4&50&51.0&29.5 \\
Oct. &95&4&42&46.1 &22.8 &95&7&35&38.5&20.7 \\
Nov. &92&11&43&43.2&18.1 &91&9&35&38.6&18.6 \\
Dec. &88&8&41&44.0&18.5 &86&8&33&38.0 &18.1 \\
Total &100&3&49&52.2&23.2&100&3&39&43.6&23.7\\
\hline
\end{tabular}
\end{table*}

\begin{figure}
   \centering
  \includegraphics[width=12.5cm, angle=0]{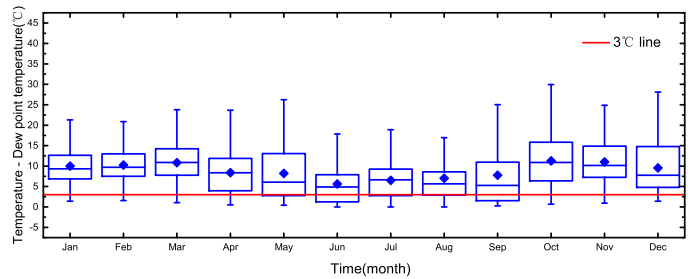}
   \caption{Boxplot of every month differences between temperature and dew point temperature over nighttime period from 2017 August to 2019 July. Each box represents the values in the range of 25$\%$ to 75$\%$ and vertical line represents the values 1$\%$ to 99$\%$,, the diamonds and horizontal lines inside every box represent mean and median values respectively. Red line represents the upper limit of 3$^{\circ}$.}
   \label{fig:xj_11}
\end{figure}

It is important to know the percentage of observational time in which condensation may occur. For this purpose, we calculate the dew point temperature during nighttime by theoretical model\citep{2016PASP..128c5004T},and give the statistics of every month from August 2017 to July 2019 in Figure~\ref{fig:xj_11}. Usually the instrumentation is a few degrees lower than the air temperature, so we set 3$^{\circ}$$C$ as the upper limit for the difference between air temperature and dew point temperature. In Figure~\ref{fig:xj_11}, the red line represents the lower limit of 3$^{\circ}$$C$, each box represents the values in the range of 25$\%$ to 75$\%$ and vertical line represents the values 1$\%$ to 99$\%$, the diamonds and horizontal lines inside every box represent mean and median values respectively. From Figure~\ref{fig:xj_11} we can see that the median values of differences between temperature and dew point temperature in autumn is more closer to the red line than other seasons which indicates that condensation may occur frequently in autumn nights.

\subsection{Barometric Pressure}
\begin{figure}
   \centering
  \includegraphics[width=13.5cm, angle=0]{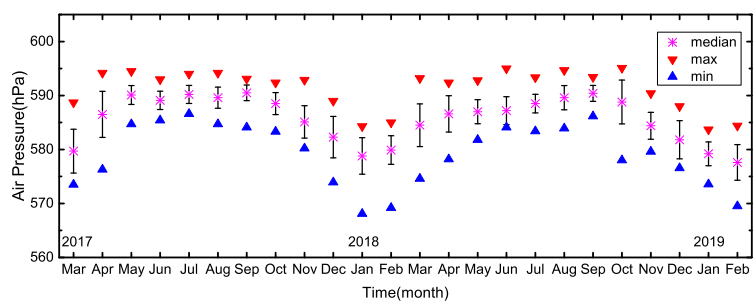}
   \caption{Monthly statistic of the air pressure from March 2017 to February 2019.}
   \label{fig:xj_12}
\end{figure}

Figure~\ref{fig:xj_12} shows the maximum, minimum, median and standard deviation values of air pressure data every month from 2017 March to 2019 February. A clear positive correlation with air temperature can be observed in Figure~\ref{fig:xj_12}, it indicates obvious plateau continental climate characteristic which is higher and with smaller variation range in warm seasons than in cool seasons.

\begin{figure}
   \centering
  \includegraphics[width=12.5cm, angle=0]{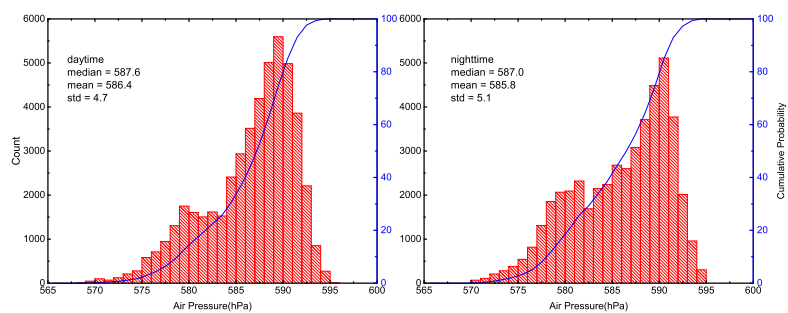}
   \caption{Air pressure histograms and cumulative distributions on daytime (left) and nighttime (right).}
   \label{fig:xj_13}
\end{figure}

The monthly maximum, minimum, median, mean and standard deviation values for daytime and nighttime are summarized in Table~\ref{Tab:xj_4}. The lowest air pressure (568.1 $hPa$) during the first year (March 2017 to February 2018) was exhibited in January 2018, whereas the highest air pressure (594.9 $hPa$) in May 2017. The lowest (569.5 $hPa$) and highest (595.1 $hPa$) air pressure during the second year (March 2018 to February 2019) have been recorded in February 2019 and October 2018 respectively.
Statistics on daytime and nighttime data are represented in Figure~\ref{fig:xj_13}. The median value during daytime is 587.6 $hPa$, and the mean value is 586.4 $hPa$. The median value during nighttime is 587 $hPa$, and the mean value is 585.8 $hPa$. The air pressure is slightly lower over nighttime period.

We analyzed the air pressure in order to see if the site is dominated by high pressure which would imply prevailing stable good weather\citep{2012MNRAS.422.2262R}.
The theoretically expected air pressure can be calculated by the Equation.\ref{eq1} (US Standard Atmosphere mode\citep{2016PASP..128c5004T}), where $P_{0}$=1013.25 $hPa$ is the sea level standard atmospheric pressure, and $T_{0}$=288.15 $K$ (15$^{\circ}$$C$) is the sea level standard temperature, $h$ represents the altitude (4526 $m$) of our site.

\begin{center}
\begin{equation}
\label{eq1}
P=P_{0}\cdot(1-0.0065\cdot\frac{h}{T_{0}})^{3/5}
\end{equation}
\end{center}

\begin{figure}
   \centering
  \includegraphics[width=13.5cm, angle=0]{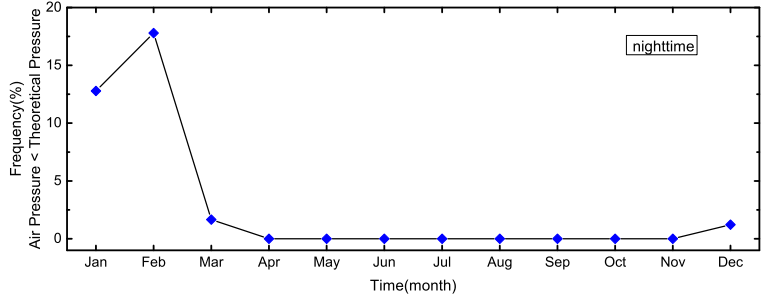}
   \caption{Monthly distributions of percentages for differences between air pressure and theoretical pressure whose values less than 0 $hPa$ over nighttime period.}
   \label{fig:xj_14}
\end{figure}

When the air pressure is lower than the theoretically expected value unstable weather may occur. We have computed the theoretically expected values for each time point during nighttime and compare them with the actual values to find out the frequency of unstable weather conditions monthly, the result is shown in Figure~\ref{fig:xj_14}. The percentage values in most of the months are roughly equal to 0$\%$, but an obvious increasing in spring from January to March, especially in February 17.8$\%$ of the nighttime period dominated by low pressure.

\begin{table*}
\centering
\caption{Monthly Statistics on Daytime and Nighttime Data of Air Pressure.}
\label{Tab:xj_4}
\begin{tabular}{lllllllllll}
\hline
\multirow{2}{*}{Month}
        & \multicolumn{5}{c}{Nighttime Air Pressure} &\multicolumn{5}{c}{Daytime Air Pressure}\\
         \cmidrule(l){2-6} \cmidrule(l){7-11}
   &Max &Min  &Median  &Mean  &Std &Max &Min  &Median  &Mean  &Std \\
\hline
Jan. &586.5& 570.1& 579.4& 578.6& 2.9& 584.3& 568.1& 579.1& 578.9& 2.8\\
Feb. &584.9& 570.1& 579.0& 578.5& 3.0& 585.0& 569.2& 579.1& 578.1& 3.2\\
Mar. &593.1& 574.4& 583.8& 583.6& 3.9& 593.2& 573.5& 583.8& 583.8& 4.3\\
Apr. &593.3& 576.4& 586.8& 585.9& 3.7& 594.2& 576.3& 586.5& 586.3& 3.8\\
May. &594.6 &582.9 &590.0 &588.9 &2.3 &594.5& 581.8&589.3&589.4& 2.5\\
Jun. &594.4 &585.3& 589.1& 588.6& 2.2& 595.0& 584.1& 589.3& 589.4 &2.3\\
Jul. &594.1 &584.7& 589.8& 589.3& 1.8& 594.0& 583.4& 589.3& 589.9 &2.0\\
Aug. &594.9 &585.4& 590.1& 589.6& 1.9& 594.7& 583.9& 589.6& 590.2 &2.1\\
Sep. &593.7 &583.9& 590.7& 590.3& 1.4& 593.4& 584.1& 590.4& 590.7& 1.5\\
Oct. &592.1 &578.0& 588.9& 588.2& 2.8& 595.1& 574.0& 588.6& 588.6& 3.2\\
Nov. &589.2 &580.1& 584.8& 585.3& 2.7& 592.9& 579.6& 584.8& 585.2& 2.8\\
Dec. &589.2 &574.3& 582.1& 582.3& 3.3& 589.0& 573.9& 582.0& 582.3& 3.4\\
Total &594.9 &570.1& 587.0& 585.8& 5.1& 595.1& 568.1& 587.6& 586.4&4.7\\
\hline
\end{tabular}
\end{table*}

\subsection{Wind Speed and Direction}
\begin{figure}
   \centering
  \includegraphics[width=12.5cm, angle=0]{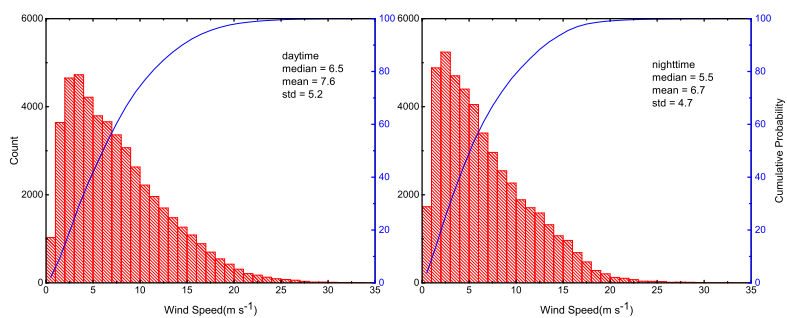}
   \caption{Wind speed histograms and cumulative distributions on daytime (left) and nighttime (right).}
   \label{fig:xj_15}
\end{figure}

Strong wind or gusts represent a serious hazard for the instruments\citep{2016PASP..128c5004T}. Figure~\ref{fig:xj_15} shows the distributions and cumulative statistics for the wind speed during daytime, and nighttime periods. The median values are 6.5 $ms$$^{-1}$ and 5.5 $ms$$^{-1}$ for daytime and nighttime respectively. Consider 15 $ms$$^{-1}$ as the limit beyond which the telescope should be turned off and brought to the parking position, the percentage beyond this limit during nighttime is 6.6$\%$.

High relative humidity values mentioned in section 4.2 and strong wind both are constraint for telescopes¡¯ operation so it is essential to estimate the combined action of them. We calculated the percentage that either relative humidity higher than the threshold 90$\%$ or wind speed stronger than 15 $ms$$^{-1}$  during nighttime periods and the result is 13$\%$, which is roughly equal to the sum of 6.8$\%$ caused by high relative humidity and 6.6$\%$ caused by strong wind. It reflects the relatively independency between high relative humidity and strong wind. Normally high relative humidity comes with precipitation and cloudy but strong wind may occur in clear night, so strong wind has to be taken into account besides cloudy amount when evaluating the annual observable time.

\begin{figure}
   \centering
  \includegraphics[width=12.5cm, angle=0]{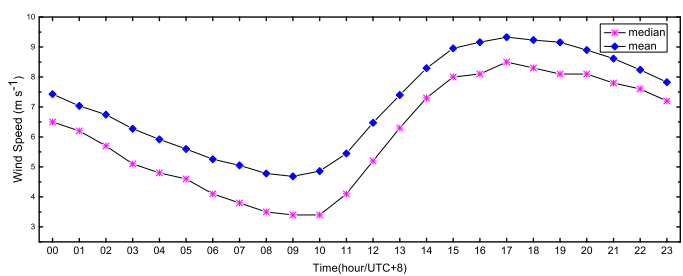}
   \caption{Hourly wind speed mean, median values.}
   \label{fig:xj_16}
\end{figure}

The day trend of wind speed is shown in Figure~\ref{fig:xj_16} where hourly medians and means are computed after merging different days. A clear time dependency can be seen that the wind speed usually starts to increase at noon and reaches the peak before nightfall, then decreases gradually during the night. Relatively low wind speed in nighttime is beneficial to astronomical observation.
\begin{figure}
   \centering
  \includegraphics[width=12.5cm, angle=0]{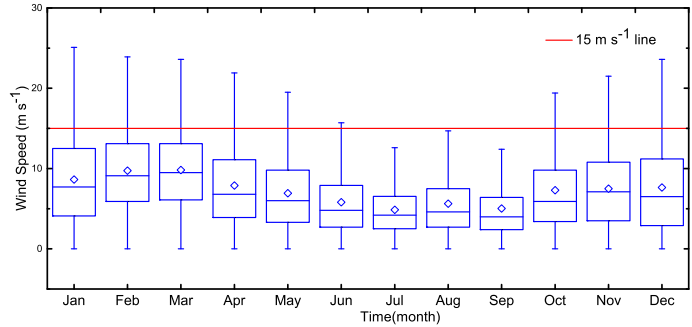}
   \caption{Boxplot of monthly wind speed over nighttime period. Each box represents the values in the range of 25$\%$ to 75$\%$ and vertical line represents the values 1$\%$ to 99$\%$, the diamonds and horizontal lines inside every box represent mean and median values respectively. Red line represents the upper limit of 15 $ms$$^{-1}$.}
   \label{fig:xj_17}
\end{figure}

Monthly statistics for the nighttime measurements of wind speed are shown in Figure~\ref{fig:xj_17}, each box represents values in the range of 25$\%$ to 75$\%$ and vertical line represents the values 1$\%$ to 99$\%$, the diamonds and horizontal lines inside every box represent mean and median values respectively. Red line in Figure~\ref{fig:xj_17} represents the upper limit of 15 $ms$$^{-1}$. It can be noticed that the wind speed is higher in cold seasons, especially in these three months from January to March the median of nighttime wind speed higher than 7 $ms$$^{-1}$. It is correspond to the frequency of low barometric pressure as seen in Figure~\ref{fig:xj_14}. More detailed statistics of wind speed over daytime and nighttime periods can be seen in Table~\ref{Tab:xj_5}.

\begin{figure}
   \centering
  \includegraphics[width=12.5cm, angle=0]{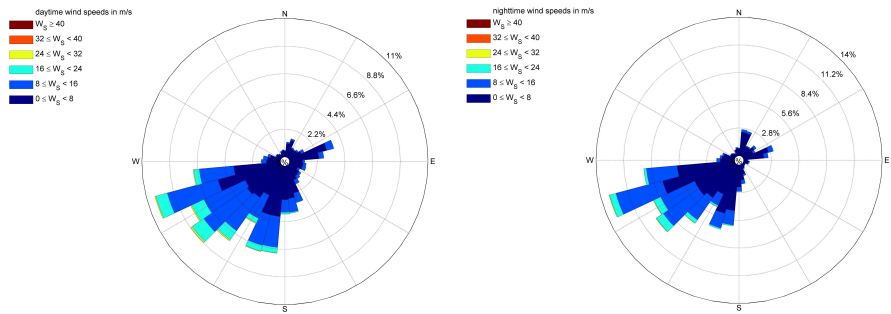}
   \caption{Wind rose density over daytime (left) and nighttime (right) period.}
   \label{fig:xj_18}
\end{figure}

The strength and direction of ground winds define the airflow conditions which influences the ground- layer turbulence\citep{2006PASP..118.1048G}).
A wind rose diagram is one gives the percentage of the time in which the wind blows from each direction, it also indicates the strength of the wind velocity in the prevailing wind directions. Figures~\ref{fig:xj_18} shows the wind rose for daytime and nighttime over the period from 12$^{st}$ July 2017 to 28$^{st}$ February 2018, from which we can clearly see that the prevailing wind direction is from southwest at all times of the year. Stable wind direction means stable airflow conditions, which will provide good observation conditions for astronomical observations. It also indicates that the wind direction is little influenced by local topography.

\begin{table*}
\centering
\caption{Monthly Statistics on Daytime and Nighttime Data of Wind Speed.}
\label{Tab:xj_5}
\begin{tabular}{lllllllll}
\hline
\multirow{2}{*}{Month}
        & \multicolumn{4}{c}{Nighttime Wind Speed} &\multicolumn{4}{c}{Daytime Wind Speed}\\
         \cmidrule(l){2-5} \cmidrule(l){6-9}
   &Max  &Median  &Mean  &Std &Max  &Median  &Mean  &Std \\
\hline
Jan. &29.0&7.4&8.3&5.4&32.8&8.0&9.1&5.7\\
Feb. &29.3 	&8.6 	&9.0 	&4.6 		&35.3 	&10.2 	&10.6 	&5.8 \\
Mar. &23.7 	&8.5 	&8.8 	&4.4 		&29.4 	&10.6 	&10.9 	&5.3 \\
Apr. &27.8 	&6.0 	&7.2 	&4.8 		&30.9 	&7.5 	&8.4 	&5.2 \\
May. &22.5 	&5.6 	&6.6 	&4.3 		&25.7 	&6.3 	&7.2 	&4.5 \\
Jun. &20.7 	&3.9 	&4.8 	&3.5 		&30.6 	&5.5 	&6.4 	&4.3 \\
Jul. &16.3 	&4.0 	&4.5 	&2.8 		&23.3 	&4.3 	&5.1 	&3.4 \\
Aug. &21.4 	&4.3 	&4.8 	&3.2 		&25.5 	&5.0 	&6.2 	&4.4 \\
Sep. &32.9 	&3.7 	&4.5 	&3.9 		&27.1 	&4.2 	&5.5 	&4.2 \\
Oct. &34.9 	&5.0 	&6.5 	&5.0 		&36.6 	&6.9 	&8.2 	&5.6 \\
Nov. &19.8 	&6.3 	&6.9 	&4.4 		&24.0 	&8.1 	&8.3 	&4.7 \\
Dec. &33.2 	&5.8 	&7.1 	&5.3 		&37.1 	&7.6 	&8.6 	&6.0 \\
Total &34.9	&5.5 	&6.7 	&4.7 		&37.1	&6.5	&7.6  	&5.2\\
\hline
\end{tabular}
\end{table*}
\subsection{Sky Brightness}
\begin{figure}
   \centering
  \includegraphics[width=12.5cm, angle=0]{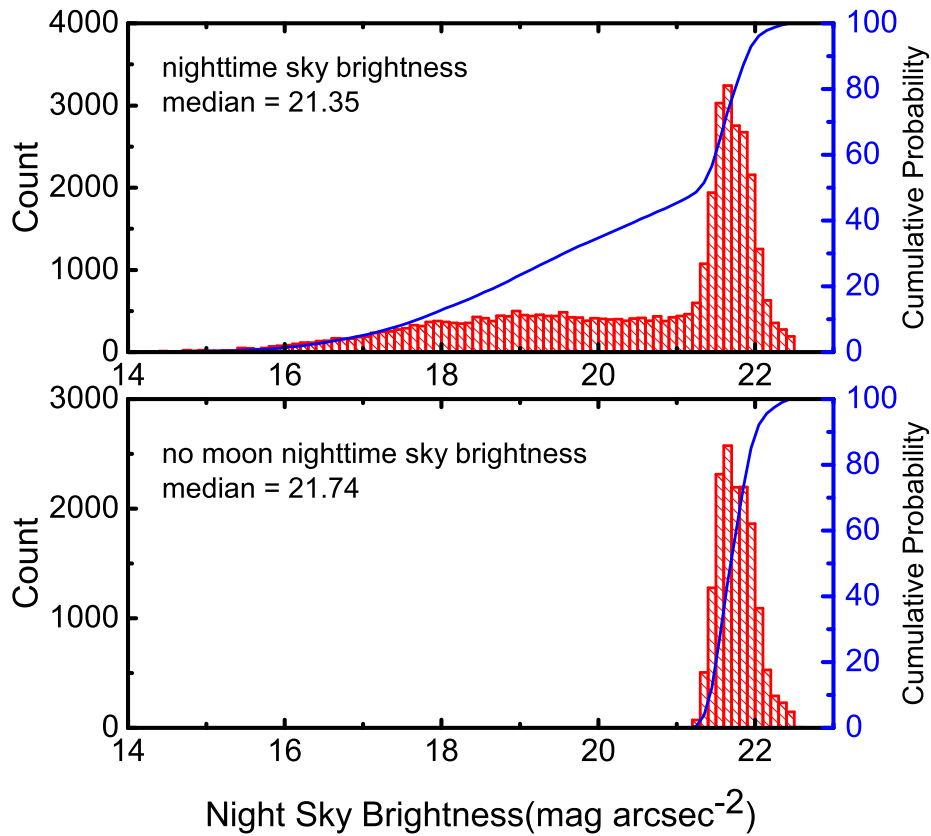}
   \caption{ Distribution and cumulative statistics of the night sky brightness (above) and no moon nighttime (below).}
   \label{fig:xj_19}
\end{figure}

Our site was equipped a commercial device SBM (Sky Background meter) developed by the Unihedron Company for sky brightness monitoring from May 24$^{th}$, 2017 to February 28$^{th}$, 2018. The Full Width Half Maximum (FWHM) of the angular sensitivity is about 20$^{\circ}$? in the zenith direction and the SBM data are sampled every minute. The histogram above of Figure~\ref{fig:xj_19} represents distribution and cumulative statistics of the night sky brightness. The median value during nighttime is 21.35 $mag$ $arcsec$$^{-2}$ and 43.4$\%$ of the values are concentrated between 21.5 and 22.3 $mag$ $arcsec$$^{-2}$. To remove the influence of moonlight, we show the sky brightness distribution during nighttime without moon (nighttime with the moon height below -6$^{\circ}$) below of Figure~\ref{fig:xj_19}, the median value is 21.74 $mag$ $arcsec$$^{-2}$.

\section{Summary and Conclusion}
A site evaluation has been initiated for the identification of the best location in China where to host the LOT project, Muztagh-ata site is one of the three candidates.We have studied the astro-climate conditions of the Muztagh-ata site base on the statistical analysis of parameters which effect on astronomical observation directly include air temperature, relative humidity, dew point temperature, barometric pressure, wind speed, wind direction, sky brightness and showed the results in this paper.

We are aware of that two years of meteorological monitoring cannot be sufficient for a full site characterization, but the main results as follow:

1.The mean value of nighttime air temperature is -6.2 $^{\circ}$$C$ while -1.0 $^{\circ}$$C$ for daytime. The maximum and minimum value during the two-year site-testing period are 17.8 $^{\circ}$$C$ and -26.5 $^{\circ}$$C$. Temperature difference shows strong seasonal dependency, greater in warm seasons and smaller in cold seasons.The annual average temperature of the two years is -2.4 $^{\circ}$$C$ (March 2017 to February 2018) and -5.1 $^{\circ}$$C$ (March 2018 to February 2019) respectively. Comparing with the result of -6 $^{\circ}$$C$ given by the local meteorological department, 2017 is significantly warmer, more precipitation may occur in 2017 than normal years. It is meaningful for analyzing the difference of cloudy amount between these two years.

2.The median of relative humidity data for nighttime from August 2017 to July 2019 is 49$\%$ and 39$\%$ for daytime. The percentage of the nighttime in which the relative humidity values higher than 90$\%$ is 8.8$\%$ and under these conditions the telescope should stopped. 59$\%$ nighttime data concentrate in the range from 20$\%$ to 60$\%$, it is mainly due to the surrounding glaciers which play a certain role in stabilizing the relative humidity during the night.

3.In order to evaluate the risk of condensation we have computed the monthly percentages for the difference values between dew point temperature and air temperature smaller than 3$^{\circ}$$C$ during night. The monthly median values are more closer to 3$^{\circ}$$C$ in late summer and autumn than other seasons, it is mainly related to the more precipitation in these months every year at Muztagh-ata site.

4.The analysis of the barometric pressure data shows typical plateau continental climate characteristics, lower and with wider range in cool seasons. A pronounced seasonal dependence was represented in the monthly frequency of the nighttime in which the air pressure value is lower than the theoretically expected value that means unstable weather conditions caused by it. The percentages are much higher in spring months especially in February, during which it reaches 17.8$\%$, than other months in which the rate roughly equal to 0$\%$.

5.The median value of nighttime wind speed is 5.5 $ms$$^{-1}$ and 6.5 $ms$$^{-1}$ for daytime. For 6.6$\%$ of the nighttime the wind speed is greater than 15 $ms$$^{-1}$. We represent the wind speed distribution per hour, it can be clearly observed that the wind velocity usually increases in daytime and decreases in nighttime, good for nighttime observation. A strong correlation between wind speed distribution and low air pressure frequency was represented. The months whose frequencies of low air pressure are relatively high, with more occurrence of strong wind.

6.The prevailing wind directions are derived from the wind rose plot. The wind comes from southwest almost both in daytime and nighttime, which could provide stable ground layer turbulence.

7.The median value of the sky brightness at Muztagh-ata site measurements is 21.35 $mag$ $arcsec$$^{-2}$ and 21.74 $mag$ $arcsec$$^{-2}$ for no moon nighttime. Through the analysis of images from all sky camera we found that the light pollution mainly due to the lighting for the debugging of monitoring equipment during nights and the village or road surrounding our site have little influence on the results.

\begin{acknowledgements}
This work is support by the program of the National Nature Science Foundation of China:11873081, 11603065 and  the  Operation,  Maintenance  and  Upgrading  Fund  for Astronomical Telescopes  and  Facility  Instruments,  budgeted  from  the  Ministry  of Finance of China (MOF) and administrated by the Chinese Academy of Sciences (CAS).
\end{acknowledgements}


\label{lastpage}

\end{document}